# CRISTAL : A Practical Study in Designing Systems to Cope with Change

Andrew Branson[1], Richard McClatchey[1*], Jean-Marie Le Goff[2] and Jetendr Shamdasani[1]

[1] Centre for Complex Cooperative Systems, University of the West of England, Coldharbour Lane, Bristol BS16 1QY, UK. Tel: +44 (0)117 328 3176. FAX: +44 (0) 117 328 2734

[2] CERN, 1217 Geneva 23, Switzerland. Tel: +41 (0) 22 767 6159. FAX: +41 22 (0) 22 767 6555

Email:{Andrew.Branson,Richard.McClatchey,Jean-Marie.Le.Goff,Jetendr.Shamdasani}@cern.ch

*Corresponding Author : Professor Richard McClatchey

## Abstract

Software engineers frequently face the challenge of developing systems whose requirements are likely to change in order to adapt to organizational reconfigurations or other external pressures. Evolving requirements present difficulties, especially in environments in which business agility demands shorter development times and responsive prototyping. This paper uses a study from CERN in Geneva to address these research questions by employing a 'description-driven' approach that is responsive to changes in user requirements and that facilitates dynamic system reconfiguration. The study describes how handling descriptions of objects in practice alongside their instances (making the objects self-describing) can mediate the effects of evolving user requirements on system development. This paper reports on and draws lessons from the practical use of a description-driven system over time. It also identifies lessons that can be learned from adopting such a self-describing description-driven approach in future software development.

## 1. Research Problem and Related Work

Many organizations operate in environments which dictate change at an unforeseeable rate and force evolution in system requirements; in these cases system development does not have a definitive end point, rather it continues in a mutually constitutive cycle with the organization and its requirements. Moreover, if that period of design is of such duration that the technology may well evolve, then the design process becomes considerably more difficult. In these circumstances the design must also need to be able to evolve in order to react to changing technologies and to ensure traceability between the fluid design and the evolving system specification. In other words extended design periods necessitate adaptable design specifications to cope with changes in requirements over a temporal dimension.

Efforts to tackle the problem of coping with design evolution have included design versioning [1], 'active' object models [2],[3], capture and exploitation of so-called mesodata [4], and schema versioning [5]. However, none of these approaches enables the design of an existing running system to be changed responsively and for those changes to be reflected in a new running version of that design. Such functionality is becoming increasingly important with the advent of reduced design to production times, especially in the business-to-business sector where time to market has become a critical driver. Businesses now expect systems to be agile in nature, to be able to cope with heterogeneity between systems and to be responsive to changes in user requirements so that they can evolve over time as the user needs change. Systems based around ERP (Enterprise Resource Planning [6]) have emerged from vendors such as Sage, Lawson, ORACLE and Microsoft which capture business functions in a centralized database but these systems have been unable to adapt to a changing business environment due to their static nature. The research challenge here is what form of design approach can be followed to ensure systems can be "future-proofed" to cope with new requirements as they emerge. This paper investigates how far object-oriented design can be pursued to this end.

Business Process Management (BPM [7]) tools have been proposed by the likes of SAP, ORACLE and IBM to handle the problem of harmonizing different cooperating systems and integrating functionality. The object of these orchestration tools is to allow decision-makers and technical teams to collaborate to define and optimize business processes by using a single tool (sometime called a *process orchestration tool*) to coordinate activities across multiple processes. However these tools often do not have the ability to enable dynamic design changes to be reflected into production-ready systems. They cannot reconcile the deterministic approach required to establish a smooth-running production system with the ability to handle the emergence of unforeseen variables and the ubiquitous changes to design that are characteristic of 'real-world' operation. Few BPM solutions offer the possibility to act on the parameters and environment of the process while it is being carried out. Even fewer are able to intervene in executing processes, eliminating or skipping steps along the way or adding new steps to replace or supplement existing execution. What is needed are novel approaches that allow systems to reconfigure dynamically, so that designs can be versioned and rolled out into production to sit seamlessly alongside production systems without system designer action. It is this research question that is tackled in this paper through addressing the demands for traceability in a very demanding environment and delivering a fully functional system which is then evaluated in practice.

In this paper we report on a practical study that investigates how we can design an object-based system to cope with evolving user requirements and facilitate dynamic system configuration. We outline an approach based on object-oriented design for providing support for design evolution by implementing what we term a 'description-driven' system. That is one in which descriptions of designs are captured alongside the data they describe in a database and that can be instantiated responsively for systems that require round the clock operation. Our approach involves identifying and abstracting the elements (such as business objects, processes, lifecycles, and agents) crucial to the system under design and creating high-level descriptions of these elements which are stored, dynamically modified and managed separately from their instances, but are treated in the same way as are their instances. The effect of this is not only to enable system evolution with changing requirements but also to simplify the management of the system in operation.

Model-driven software development [8] has been proposed as one method that can provide solutions for systems whose requirements evolve over time. There have been a number of model-driven systems that have emerged in the last decade that go some way to providing such support, examples of these include ERP/BPM systems such as ADempiere [9], OpenBravo [10], Apache OFBiz [11] and Compiere [12]. All of these systems allow users to reformulate workflows and data structures on-the-fly, and provide varying levels to control the management of changes, but they all implement their meta-models entirely separately from their instances. Our CRISTAL approach extends the model-driven concept in both the descriptions and the instances of those descriptions being implemented as objects and being maintained using the same software. Even though workflow descriptions and instance implementations are different, the manner in which they are stored and are related to each other is the same in CRISTAL. This approach is similar to the distinction between Classes and Objects in the original definition of object oriented principles [13]. We have followed those fundamental principles in CRISTAL to ensure that we can provide the level of flexibility and maintainability that object orientation can enable to facilitate system evolution.

This paper describes the motivation for this research in terms of a practical example based around large-scale engineering. The description-driven design philosophy is outlined in Section 3 before discussion of the CRISTAL architecture is described and how a prototype description-driven system was built and evolved into a production system in Section 4. The system was evaluated over an eight year period and the lessons learned from running that production system are presented in Section 5 and 6 before conclusions are drawn on how such systems might further evolve and be used in other domains.

## 2. A Practical Example of Handling Evolving Requirements

This research work was motivated by the research questions that arise when handling the evolving requirements of users that are concurrently working over extended timescales, in this case on the design of a large scientific experiment. These challenges although demanding are the same in principle as those for any system that has been designed and operated over long durations of time. The example studied was that of the Large Hadron Collider (LHC) accelerator project at CERN [14], the European Centre for Nuclear Research, which after more than a decade of research and development reached its final phase of construction and testing in 2009. The Compact Muon Solenoid [15] is one of the four main experiments of the LHC; it contains a number of complex scientific detectors for identifying the constituent particles emerging from proton-proton collisions at the LHC. Amongst these is the so-called Electromagnetic Calorimeter (ECAL, comprising an enveloping array of about 70,000 individual single crystals of lead tungstate ($PbWO_4$) together with their associated fast front-end and readout. Its design and construction were carried out by a team of 1500 engineers and scientists in over 150 institutions distributed worldwide between 1995 and 2008 and form the subject of this research study. Its requirements are outlined in the following paragraphs.

The construction process of detectors for the LHC experiments is long scale, heavily constrained by resource availability, and evolves with time to account for improvements guided by research from iterative prototyping. As a consequence, changes in detector component design need to be tracked and quickly reflected in the construction process. Each constituent part of each detector needed its physical characteristics to be measured, producing Terabytes of data and had to be tested locally, prior to its shipping and assembly and integration in the experimental area at CERN. Furthermore each part must be labelled and located in the final assembly. Much of the data collected during this phase is needed not only to construct the detector, but for the later calibration and operation. Thus any system for managing this data must provide traceability (i.e. provenance) of composite parts and processes and be evolvable over time.

The huge cost and extended time-scales of the ECAL project created circumstances which, though not unique, were extreme in nature and presented real demands in terms of requirements:

- Component parts were very expensive. Wastage had to be minimised, particularly since the Compact Muon Solenoid was a public-funded research experiment. Therefore prototypes often needed to be recycled or retrofitted to be incorporated into the final assembly and new and improved construction techniques, whose versions required logging, were developed iteratively to keep costs under control.

- The Compact Muon Solenoid design and construction was a state-of-the-art research project where new techniques for electronics and for crystal growth, characterisation and usage were being discovered as the project progressed. Lead tungstate is brittle, expensive and difficult in the extrusion of 30cm single crystals and great care was required in hits handling.

- As a consequence, untested designs and technological progress over the long duration of construction extended the design phase into the production phase to a large extent so that multiple versions of component parts and measured characteristics had to be tracked concurrently, but ultimately had to be seamlessly accessed as a single source of data at any point in the production phase.

- Production of the crystals and their testing was distributed across many sites in Europe and Asia. The data had to follow the individual crystals, their electronics and their assemblage and be traceable over the 10 years of design and production and for up to 15 years of running that followed from 2008.

All of this meant that the initial logic for ECAL construction would change considerably throughout the project, while the data needed to be gathered, managed and curated for future use. The specification of the ECAL took more than a decade during which time many fundamental decisions impacting its design, detector components' physical characteristics and data formats were made. Therefore, any database system used to track those changes had to be able to handle frequent changes and thus had to be very flexible and to evolve in design.

Since at the time no existing ERP system could address the demanding requirements of CMS, a research project, entitled CRISTAL ([16], [17]) was established to develop a support system for ECAL construction; this project followed as far as was practically possible, pure object-oriented principles, languages and design techniques, to facilitate the management of the data collected during construction of the ECAL. The data sources were all the physical characteristics of detector components, which were required by the physicists for activities such as detector construction and maintenance. This included characteristics such as physical dimensions, light yield, transmission, doping etc. of the lead tungstate crystals. CRISTAL is a distributed product data and workflow management system which has a repository with a multi-layered architecture for component abstraction and dynamic modelling for the design of the objects and components of the system. These techniques were employed to handle the complexity of such a data-intensive system and to provide the flexibility to adapt to the changing production scenarios typical of any research production system.

In summary, the CRISTAL project aimed to implement a distributed engineering information management system, which orchestrated the construction process of ECAL. Its demanding objectives :

- To provide an information system to control quality in the assembly of ECAL during construction;
- To monitor the global production process across distributed centres;
- To capture and store component data and testing procedures during detector construction;
- To integrate instruments measuring physical properties of detector components used to characterize them individually for optimal positioning in the final detector;
- To provide controlled, multi-user access to the production management system; and
- To provide access for Compact Muon Solenoid users to the detector engineering and calibration data computed from the physical properties collected during the execution of the assembly process.

### 3. Theory: CRISTAL Design Philosophy

Designing a system to provide product data management, workflow tracking and change management to an agreed set of user requirements is a challenging task and there have been many previous projects aimed at this (see [18]). However in a research environment, such as ECAL, where new materials and associated electronics are constantly being evaluated and developed it is simply not possible to specify at the outset of a 10-12 year effort precisely the detail of the final detector, nor the production processes for its construction. Researchers needed to track all the steps from early prototyping to completion, entailing the capture of design versions and the storage of all data associated with instantiations of those versions.

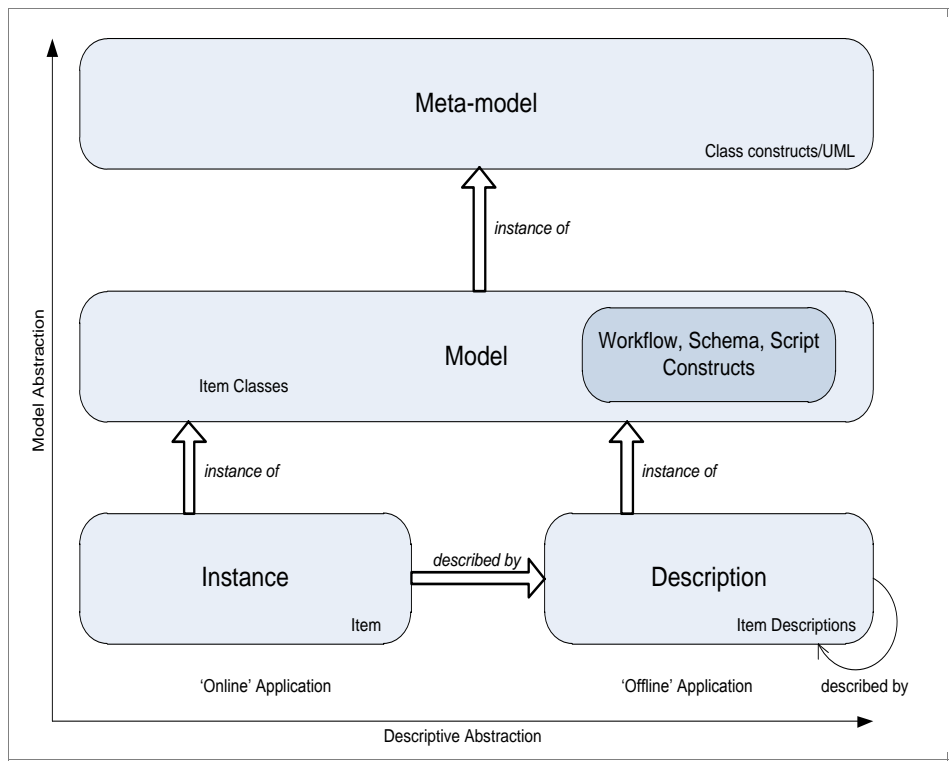

*Figure 1 - Model vs. Description in Cristal 2 (from [21])*

The CRISTAL system that has been developed to address the design process of the ECAL has been based on a so-called *description-driven* approach in which all logic and data structures have "descriptions" which can be modified and versioned online as the design of the detector changes. Research has shown that developing such meta-models can help create a flexible system offering the following - reusability, complexity handling, version handling, system evolution and interoperability (see [19] and references therein). Promotion of reuse, separation of design and implementation and reification are some further reasons for using meta-models [20]. A description-driven system (DDS) architecture, as advocated in this paper and previously in [21], is an example of a reflective meta-model architecture. DDSs make use of meta-objects to store diverse domain-specific system descriptions (such as items, processes, lifecycles, goals, agents and outcomes) which orchestrate the lifecycles of domain objects. Described as objects, the reified system descriptions of DDSs can be organized into libraries that conform with well-understood frameworks for object-oriented languages and to their adaptation for specific domains.

Following the conventions from the OMG Meta-Object Facility (MOF), Figure 1 shows how DDSs are represented as multi-layered models. The DDS two-dimensional model representations reflect the structure of the MOF with its dimensions of model abstraction [22] and descriptive abstraction [23], [24]. In DDSs the descriptions and the instantiated elements of data are stored in the database and the evolution of the design is tracked by versioning the changes in the meta-model over time. The separation of the descriptions from their instances allows them to be specified and managed and to evolve independently and asynchronously. This separation is essential in handling the complexity issues facing many computing applications; it facilitates interoperability, reusability and system evolution as there is a clear delimitation between the application's basic functionality and its representations and control. Separating descriptions from their instantiations allows new versions of defined objects to coexist with older versions. The reader is directed to previous publications ([16], [17], and [25]) for further background on DDSs.

At the outset of the CRISTAL project an object-oriented design approach was followed that enabled both ECAL parts, (with the descriptions of their specifications) and the activities, (with the descriptions of their specifications) that were carried out on the parts to be saved side-by-side but managed separately in a structured database. In this way as physicists developed their research ideas, we were able to capture each design version with those parts and activities that were processed with that version. The first CRISTAL version was driven by the requirements from the ECAL community and it implemented descriptions and data separately. This version used the Objectivity database for its backend repository as has been previously described in [16] and [17]. CRISTAL V1.0 went into production in 2000 when the first lead tungstate crystals were grown, tested and characterised on specialist automated equipment at CERN. However, due to its complex layered architecture, it suffered from performance limitations and was replaced with a second, improved, version in 2003 [26]. Essentially the initial version was not multi-threaded, and it was dependent on the object database product, Objectivity [27].

The second CRISTAL version (V2.0) had an enlarged set of requirements covering both the Compact Muon Solenoid ECAL requirements and those from commercial BPM users. This enlargement led to the generalization of the concept of 'parts', along with their descriptions, to that of 'items' which can function as either the described parts or their description objects, or indeed any domain object in the system. The items were intended to be applicable to describe any element of any domain process, and the definitions of those processes. This novel 'self-describing' version used a common design for model and instance management. This meant that the development of the descriptions, which implemented the domain application, were able to be handled in the same flexible manner that parts had been with CRISTAL V1.0. Changes could be tracked and audited, and descriptions were as portable between construction centres as the parts they were manufacturing. Other improvements on this version included multithreaded network listeners, a database switch from Objectivity to a choice of relational DBMSs, and a much reduced codebase which consequently yielded much improved performance. The second version of the CRISTAL kernel was spun off as a BPM application for industry to the Agilium company in 2003 [28]. The next section describes how CRISTAL ran in practice as an example of catering for evolving user requirements.

## 4. The Final CRISTAL Production System: Modus Operandi

CRISTAL V2.0 is an application server that abstracts all of its business objects into workflow-driven, version-controlled *'Items'* (italicised words define CRISTAL meta-objects hereafter). *Item* objects are instantiated from the Item class and are populated by Item Description objects (that are themselves also *Item* objects, see the 'described by' loop on Description in Figure 1) and are managed on-the-fly for specific user domains. At the lowest level of figure 1, *Items* have descriptions and are instantiated in the online and offline applications in CRISTAL. At the model layer, *Items* have Item classes and descriptions also use the same Item class objects. In other words they are instantiated from the same classes in the object-oriented language that is used to build the applications (in this case Java). *Items* exist to trace a *Workflow* which guides them in the storage and versioning of XML fragments.

XML is a key technology to CRISTAL, for two reasons; firstly several libraries exist in CRISTAL to marshall Java objects to XML, meaning workflows can store and version objects easily and secondly XML can be described using XML Schema, which is itself XML. Consequently, every object and piece of data in CRISTAL, from product characterization data and its definitions, to workflow, to workflow description, application logic scripts, and collection definitions are stored in the same way, as XML fragments in *Items* and are consequently simpler to manage and manipulate. To that end, *Items* contain:

- *Workflows* i.e. complete layouts of every action that can be performed on that item, connected in a directed graph that enforces the execution order of the constituent activities.
- *Activities* capture the parameters of each atomic execution step, defining what data is to be supplied and by whom. The execution is performed by agents.
- *Agents* which are either human users or mechanical/ computational agents (via an API), which then generate events.
- *Events* detail each change of state of an Activity. Completion events generate data, stored as outcomes.
- *Outcomes* are XML documents resulting from each execution (i.e. the data from completion Events), for which viewpoints arise.
- *Viewpoints* refer to particular versions of an Item's Outcome (e.g. the latest version or, in the case of descriptions, a particular version number).
- *Properties* are name/value pairs that name and type items. Properties also denormalize collected data for more efficient querying, and
- *Collections* enable items to be linked to each other.

These *Item* contents need to be defined when domain systems are modelled in CRISTAL and are, crucially, also modelled using the concept of *Items*. Description items function in exactly the same way as other *Items*; their workflows consist of activities for managing the data of the description, and also contain an instantiation activity that creates new *Items* from that data in addition to identifying information for the new *Items*, such as barcodes. The construction of the specific CRISTAL model for the domain under consideration therefore concentrates on the essential enterprise objects of the system that could be needed during its lifetime no matter from which standpoint those objects are accessed. These enterprise objects each have a creation/modification/deletion lifecycle and the CRISTAL model simply keeps track of status changes to the objects over those lifecycles. In this way it can orchestrate the execution of *Workflows* on *Items* by *Agents*, log all *Events*, *Outcomes* and *Viewpoints* and thereby capture all associated audit trail (or provenance) information associated with the domain system under study.

At a low-level, the versioning mechanism that gives provenance to the *Item* instance is the same mechanism that enables concurrent versioning in the descriptions. This means that any communication between different CRISTAL servers can transfer descriptions in exactly the same way as instances. Also dependencies can be declared as easily between abstraction layers as within them. All of these advantages arise because CRISTAL extends the original object orientation concept ideas, to more of its data model than other model-driven systems, in the same way that Java gains similar advantages from implementing classes as Class objects. This is the fundamental benefit of the CRISTAL *Item*-based design.

A disadvantage to the CRISTAL design is that the definition of 'Object' in the CRISTAL system is an *Item* which, while adhering to many core concepts of object orientation, does not follow the classic Class/Object model. This is because all Descriptions, and instances of Descriptions, are defined as *Items* in the CRISTAL model. This was deemed to be necessary to simplify the understanding of the CRISTAL design. Many developers in practice find the concepts of object orientation complex to grasp, and expertise in CRISTAL adds more to this, making it difficult to understand. This is due the large amount of terminology involved in the design of CRISTAL as well as the complexity of its concepts. This was especially challenging during development, as new personnel faced a steep learning curve before they could usefully contribute to the codebase. However, we feel that *Items* represent a return to the core values of object orientation, at a time when modern languages are becoming increasingly profligate in their implementation of them in the name of efficiency, thereby sacrificing many of the benefits that object orientation can offer.

The Compact Muon Solenoid ECAL detector was made of thousands of similar parts, all needing characterizing and assembling in an optimal configuration based on sets of detailed measurements. Every component part was registered as an *Item* in the CRISTAL database, each with its barcode as an identifier, stored as the CRISTAL *Property* 'Name'. Each part had a type, which functioned as the Item Description, and was linked to the *Workflow* definition that each instance would follow in order to collect its data and mount sub-parts. The part types also contained subtype data as *Properties* and Collection Definitions to make sure that parts were assembled in assigned positions in ECAL. All collected data and assembly information were stored as *Outcomes* attached to *Events*, so the entire history of every interaction with the application was recorded. The result was a set of *Items* representing the top level components of the detector ('super-modules') which contained five levels of substructure, all with their full production history and with all collected and calculated production data attached in the correct context (see figure 2).

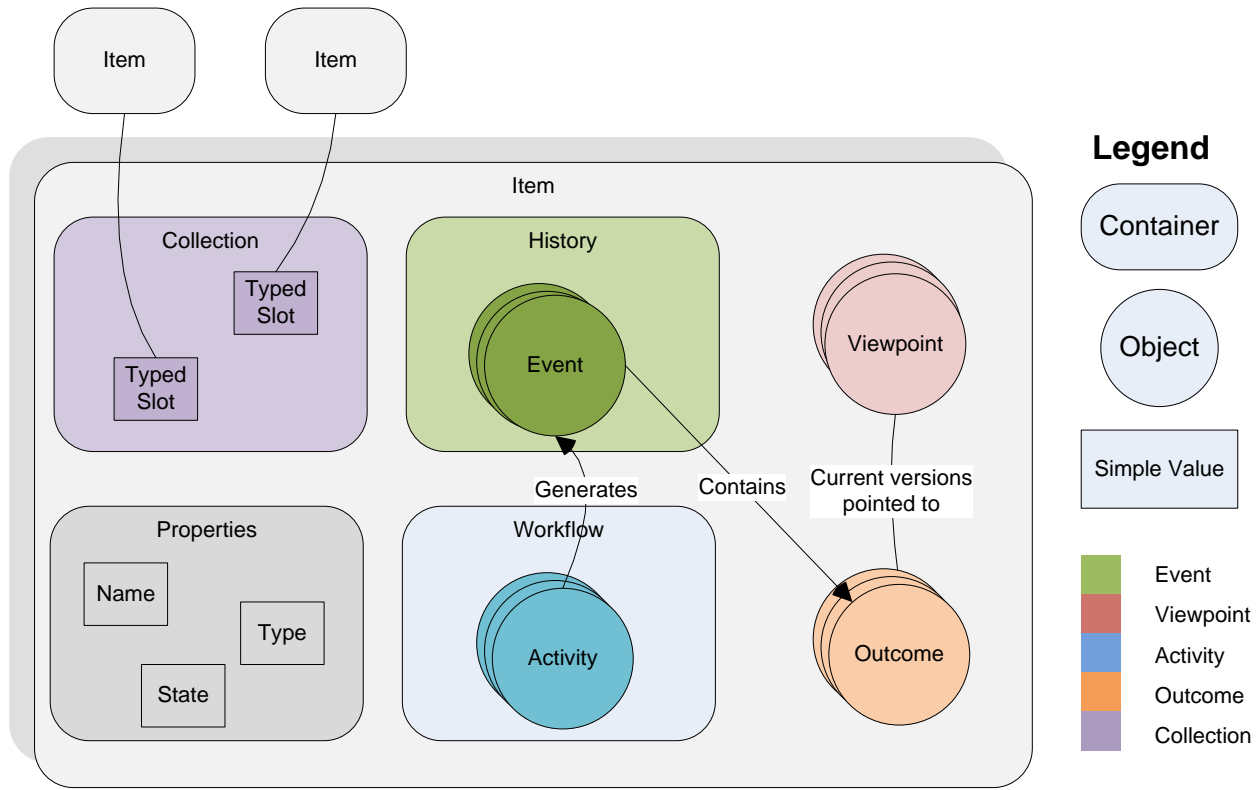

*Figure 2 - Overview of a CRISTAL Item*

On creation, an *Item* contains no *Events*, *Outcomes* or *Viewpoints*. These are explicitly generated later during execution of its workflow processes. Initially an *Item* contains *Properties* to identify it, its *Workflow* and any *Collections* it may need, with all slots empty. The initial set of *Properties* are created in the process of rendering a property input form (which is user editable) by exploiting the Property Description (itself an *Outcome* stored in the Item Description), and submitting it as an *Activity Outcome* of the description *Item*. It is this standardised and managed treatment of the workflow-driven, version controlled '*Items*' that is novel in a description-driven system and enables the flexibility of such systems.

Collection Descriptions are a special type of *Collection* that contains *Slots* that point to other Item Descriptions. The properties of the referenced Item Descriptions are copied into the slots of the new *Collection*, thereby creating *Typed Slots*, so that any instances can be checked by type against them during any future assignment. Once instantiated, the new *Collection* no longer depends on the original Collection Description, so the model is free to evolve without breaking existing instances, and the designer is free to use the type identifying *Properties* in any way that she sees fit. The *Workflow* is instantiated from a Composite Activity Description Item, which holds and versions the layout of the top level of the *Workflow*, referencing child *Activities* by name and version. These are then loaded from their Activity Descriptions, and a complete Workflow instance is created from the full Description. Scripts and Schemas are never part of the items that use them; they are loaded when the instance *Workflow* needs to use them.

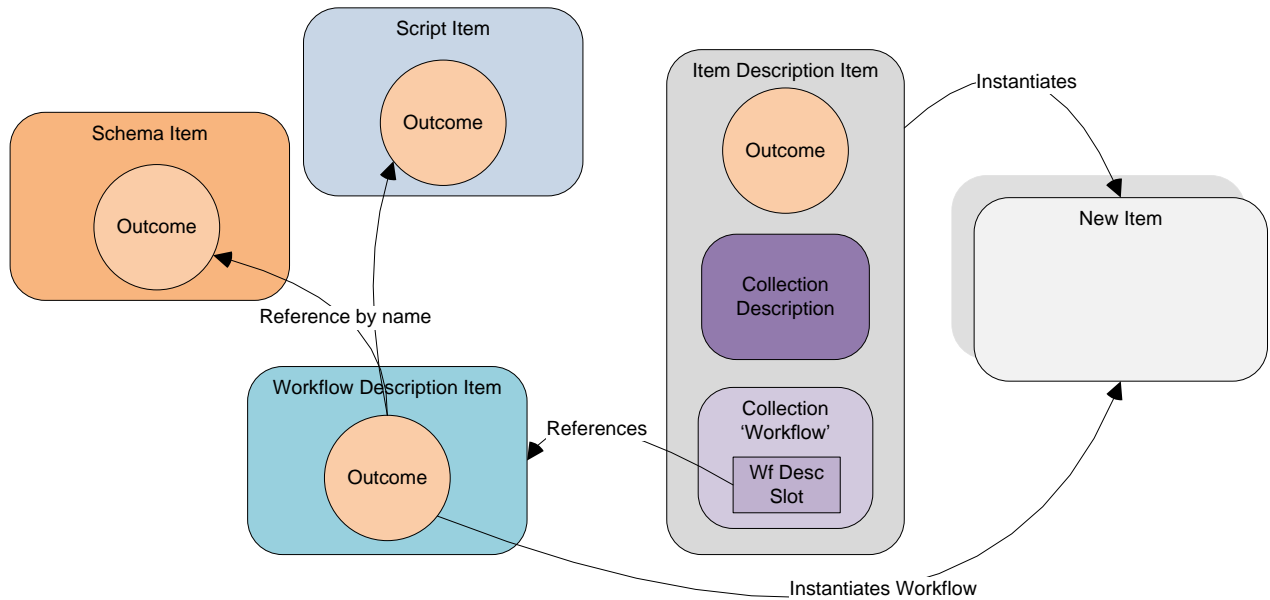

*Figure 3 – Descriptions in CRISTAL (darker colours denote descriptions of Figure 2 elements)*

Item Description *Items* hold the templates for new *Items*, and also dictate their type (see Figure 3). They are themselves also *Items* (and thus the two can be treated in the same manner), holding the description data as XML outcomes managed through *Workflow* activities. *Workflow* and *Property* descriptions are stored as XML serialized objects. Collection Descriptions are themselves *Collections*, pointing to other Item Descriptions. Outcome Descriptions contain XML Schema documents which are used to validate submitted outcomes and aid in data collection, for instance to generate data entry forms in a stock GUI for the end users. Also included in the descriptions are *Scripts* - snippets of code invoked by workflows either during a change of *Activity* state to enact consequences of the execution such as updating a *Property* or changing a *Collection*, or to assess conditional splits in the *Workflow* based on existing data. Scripts are given the full context of the *Item* and *Activity* in which they are running, so they may invoke other *Activities* (which may in turn trigger other scripts), or even navigate *Collections* to interact with other *Items*. They are stored in an XML Outcome along with declarations of their input parameters and output.

Figure 4 is a combined view of all of the components of an *Item* shown with the Descriptions used to create them. The basic functionality of CRISTAL is best illustrated with an example: using CRISTAL a user can define product types (such as Newcar spark plug) and products (such as a Newcar spark plug with serial number #123), workflows and activities (to test that the plugs work properly, and mount them into the engine). This allows products that are undergoing workflow activities to be traced and, over time, for new product types (e.g. improved Newcar spark plug) to be defined which are then instantiated as products (e.g. updated Newcar spark plug #124) and traced in parallel to pre-existing ones. The application logic is free to allow or deny the inclusion of older product versions in newer ones (e.g. to use up the old stock of spark plugs). Similarly, versions of the workflow activities can co-exist and be run on these products.

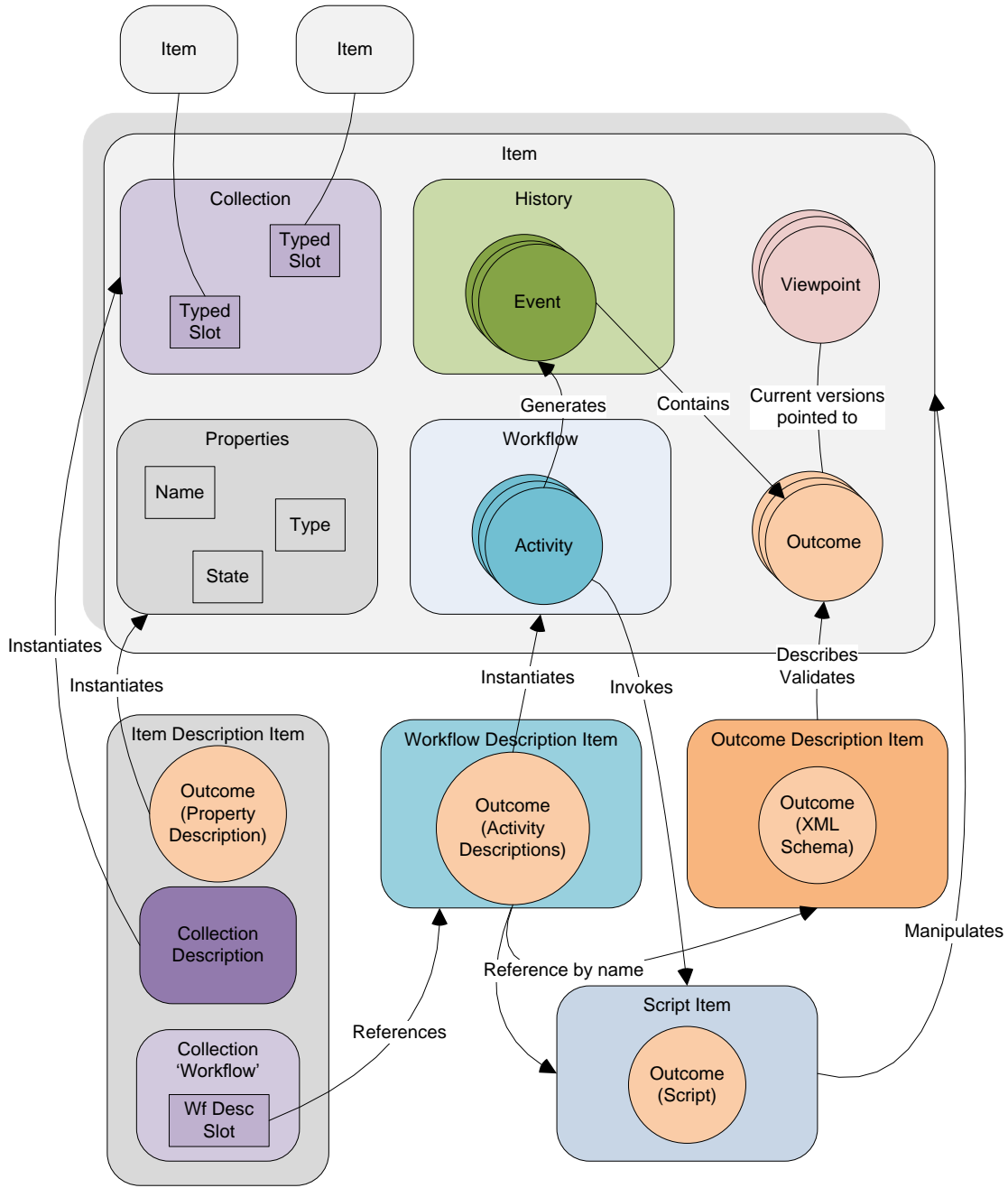

*Figure 4 - Origin of constituent objects of the CRISTAL Item, merging Figures 2 & 3. Every object above (from figure 2) has its origin below (from figure 3)*

Workflow decision Scripts are the only *Scripts* that needed to be run in the server process itself in order to determine which *Workflow* path is to be taken. These must be run in the server context, as it is the only place with the authority to make such a decision. Activity scripts are regarded as part of the *Activity* that the *Agent* is executing, therefore these are run as part of their execution procedure. Client processes do not normally have write access to the data stores - again an issue of authority to avoid distributed data transactions - so changes to the *Item*'s data are performed by invoking special 'Predefined Steps' that are attached to every domain workflow. These steps are always available for execution, though they are hidden from the workflow path. They perform such tasks as assigning/de-assigning *Items* from *Collections*, changing *Property* values, or using the current *Item* as a Description to instantiate a new one.

Most of the information used to create *Items* is stored as *Outcomes* in the description items; Description Items and thus the *Items* themselves have their own Descriptions to indicate the *Workflow* required to create those *Outcomes*. The system must bootstrap top-level descriptions of Descriptions (or meta-meta-data) at its initial run. Included in these are *Activities* and *Workflows* for managing the Descriptions, and XML schemas which detailed all of the Description types, including XML Schemas. Not obvious in the above diagram is how Collection Descriptions were created: on the description layer above they are edited in XML form as Outcomes of an 'Edit collection' *Activity*, the schema of which is compatible with the Castor mapping file which unmarshalls them during instantiation and stores them in the new instance.

In practice in the CRISTAL model of the ECAL system there are no descriptions above the first layer deployed on the production server. Descriptions needed for instantiation could be imported from files, which were stored in a version control system. The administrators had higher levels for development on their own machines, and the resulting descriptions were synchronized with a code versioning system, these were then imported into the production system. If all of the description management tools had been available at the start of the project, then the additional complexity of a separate version control system would not have been necessary; CRISTAL's description hierarchy would have done this *in situ*.

## 5. Working with Users, an Evaluation of CRISTAL in Practice

Conventional software development separates the specification phase from the construction and implementation phases. However, when the design is evolving as a result of changing user needs, the development process must be reactive and necessarily iterative in nature. The new requirements from the users need to be implemented by the developer in an incremental fashion so that the new results can be assessed and further changes to the design requested, if needed. CRISTAL allows the user to directly verify the business object workflow design, so the normal progression through implementation and testing can be short-circuited. In other words the users can visualize the overall process to be captured in terms of their own recognizable world objects; this greatly simplifies the analysis and (re-)design process. It is relatively easy for professional users to understand the workflow system in CRISTAL, and the nature of XML based data; these both can be detailed by the application maintainer sufficiently accurately in collaboration with the user or may even be drawn by an expert user directly.

The application logic that needs to be executed during the workflow will have its functionality conveniently broken down along with the activities. It is then simple to import these definitions into the system where it can be immediately tested for feedback to the users. Improvements can thereby be quickly performed online, often by modifying the workflow of one test item, which then serves as a template for the type definitions. Items subject to the improvements can co-exist with items generated earlier and prior to the improvement being made and both are accessed in a consistent, reusable and seamless manner. All this can be done without recompiling a single line of code or restarting the application server, providing significant savings in time and enables the users to work in an iterative and reactive manner that suits their research, often adopting rapid prototyping principles.

In our experience, the process of factoring the lifecycle and dataset of the new item type into activities and outcomes helps to formalize the desired functionality in the user's mind; it becomes more concrete - avoiding much of the vague and often inconclusive discussion that can accompany user requirements capture. Because it evolved from a production workflow specification driven by user requirements, rather than a desire simply to create a 'workflow programming language', CRISTAL's style of workflow correlates more closely to the users' concept of the activities required in the domain item's lifecycle. The degree of granularity can be chosen to ensure that the user feels it provides sufficient control, with the remaining potential subtasks rolled up into a single script. This is one important aspect of the novel approach adopted during CRISTAL development that has proven of benefit to its end-user community. In practice this has been verified over a period of more than 10 years use of CRISTAL at CERN and by its exploitation as the Agilium product across many different application domains (see discussion in the later conclusions section).

*5.1 From Users Requirements to a Domain Implementation*

The following phases describe the typical design process followed when the user community creates a CRISTAL system from scratch or amends an existing system to cater for evolving user requirements. It is based on an analysis of user usage patterns, requirements and capabilities that was undertaken at the outset of the CRISTAL development period and that was reviewed and amended, periodically, in consultation with the ECAL user community [29]:

1. Firstly, the design of the normal workflow of a new item type is outlined on paper with the user. As each activity is identified, any data collected during its execution is also written down. Any computation that needs to be performed is broken down into individual scripts.

2. The standard CRISTAL user interface is normally usable for executing virtually any activity, but it is rather abstract. So then the question is asked, will this suffice for this workflow, or is a simpler custom interface required? Are the end users familiar with CRISTAL or not?

3. Next, any exceptional paths (see later section) followed in the user workflows or patterns are identified. Common ones are implemented with conditional splits in the workflow. If there is too much branching, there is a risk that the user will be overwhelmed or will spend too much time dealing with the choices. Others can be omitted until they get more precisely defined through experience.

4. Once the workflow design is complete, the new types can be created in CRISTAL and one test item can be instantiated. Essential changes are implemented in the instance workflow, and the data is cleared and the workflow restarted as required. Once this is stable, the changes can be copied back to the description, and normal production can (re-)start.

One essential improvement made for V2.0 of CRISTAL was the absence of dependencies between an Item and its description, other than for type identification. For example, when a workflow is instantiated, all information required for execution is copied to the instance objects, therefore it is not necessary to refer back to the description, and the instance workflow can be modified with no ill effects. Also, this means that instance workflows are protected from inconsistencies that may be caused by modifications of their descriptions. Consequently modifications to the parent workflow description cannot be easily propagated down to pre-existing instances and such migration must be done manually or programmatically through scripts. This is a more appropriate procedure since given that a workflow cannot be guaranteed to be in a consistent state if it is modified mid-execution, so care and assessment are necessary if this modification is to be done at all.

*5.2 Integration of Users' Application Logic*

Once the activities have been broken down into a coherent workflow, any logic required to modify item components can be carried out on each activity that requires it. This provides a built-in 'top down' approach to development that is in common with other workflow based systems, although CRISTAL also allows the logic attached to each activity to contain complex code 'snippets', allowing the workflow to be simpler if the user does not require traceability to go any deeper. An example would be a single activity that empties a collection in the current item to reset assembly – although it loops through every slot freeing up the child item, the user sees it is just a single task.

CRISTAL includes a generic 'user-code' process, which can log in and execute jobs generated by activities as they change state. This will execute any scripts associated with transitions as part of the execution process, but the process can also be extended to execute other code as necessary. Thus, there are three methods available for deployment of application logic:

- As a CRISTAL script: This approach is the most desirable, since it supports versioning alongside the rest of the descriptions. Although CRISTAL supports many scripting languages through the Bean Scripting Framework, including Javascript, Python and Perl, there is still significant migration required to convert existing compiled logic.
- Using a custom user-code agent as a wrapper: This was the method used to migrate existing CRISTAL V1.0 logic, written in Java, to CRISTAL V2.0. This has the advantage of requiring little or no modification of the existing code once the wrapper is complete, but modifications require redeployment of server code.
- A mixture of the two: scripts to invoke external code.

Again, there is a choice to be made between quick fix solutions that could be more difficult to maintain and a more thorough implementation that will require little maintenance. Unlike the choice for activities, it is much harder to replace a wrapped solution with a scripted one - the new script must be developed and tested as if from scratch - but it does enable the migration to be staged over a longer time than would otherwise be available. Once the Items are designed, and their required application logic has been implemented, then the system will be ready to go into production. Users may instantiate Items from their descriptions, which will hopefully behave as the users intended. This never happens, of course, and inevitably the items will diverge from the users' needs. At this point, the user returns to the developers to request fixes and the design process iterates until the users' needs have been satisfied.

*5.3 Evolution and Maintenance Cycles*

Once the prototyping has been completed the domain application can finally go into production. From this point onwards data entered into the system must be preserved and any changes to the domain application, defined entirely in descriptions, must maintain backwards compatibility with it. The designer's work is completed, and the maintainer's begins. When an item design goes into production, the chances are that it will have to be modified due to either changing user requirements or inaccurate/unspecified initial requirements. These new requirements may manifest themselves to the domain application maintainer in a number of different ways.

Sometimes, and most conveniently, the users will present a modified workflow design directly, which after examination for possible inconsistencies can be directly implemented in the system. A user who is well versed in workflow design may even ultimately be given the rights to modify the workflow themselves. However, more often the user will request some modification by the application maintainer to the item's data or workflow, to enable such a state that is not possible with the current design. In this case, the application maintainer must assess the likelihood of this requested intervention happening again. If they think it will be a rare occurrence, then it can be quickly fixed by exceptionally manipulating the item's data directly, although care must be taken to make sure that the altered Item is left in a consistent state, limiting the number of people who can perform the fix to domain developers who have a deep knowledge of the Item design. Scripts can be written to apply an identical change to many items at once. If the original design was clearly wrong, then the workflow description can be modified for future items and existing items can be migrated if possible.

Figure 5 illustrates what happens when a user approaches the application maintainers with a particular problem that is occurring with such frequency that a system modification is required. The user requests the application maintainer to provide a rapid modification for this exceptional path in the user's workflow. Modifications start with a quick fix script which is speedily implemented and later refined to become part of the standard administrative scripts. As the modification matures from quick fix to standard script to part of the production domain the time saved exceeds the development time required and the specification becomes more concrete. The same script code used for the fix can thus eventually end up in the normal workflow path. If a particular manual intervention is used too often and is taking much of the administrator's time, he will make a reusable script, refactoring it and testing it so it can be used by others. If the user must use this script often, then the user will include it in the workflow. Demotion of the fix can occur if the workflow is eventually found to be cluttered with rarely used activities. In this way the user's modifications can be dealt with in a timely manner and the application maintainer can refine the improvement offline thereby not interfering with the production schedule.

To provide traceability, all changes of state in CRISTAL activities are themselves logged as events in the histories of items. Although it is possible to write a script in such a way as to write to the data storage directly, this method will leave no trail for later debugging and it must also be run in the server process itself (the only place with write access to the data stores). Consequently scripts interact with their items through activity execution (most usefully through the system predefined activities).

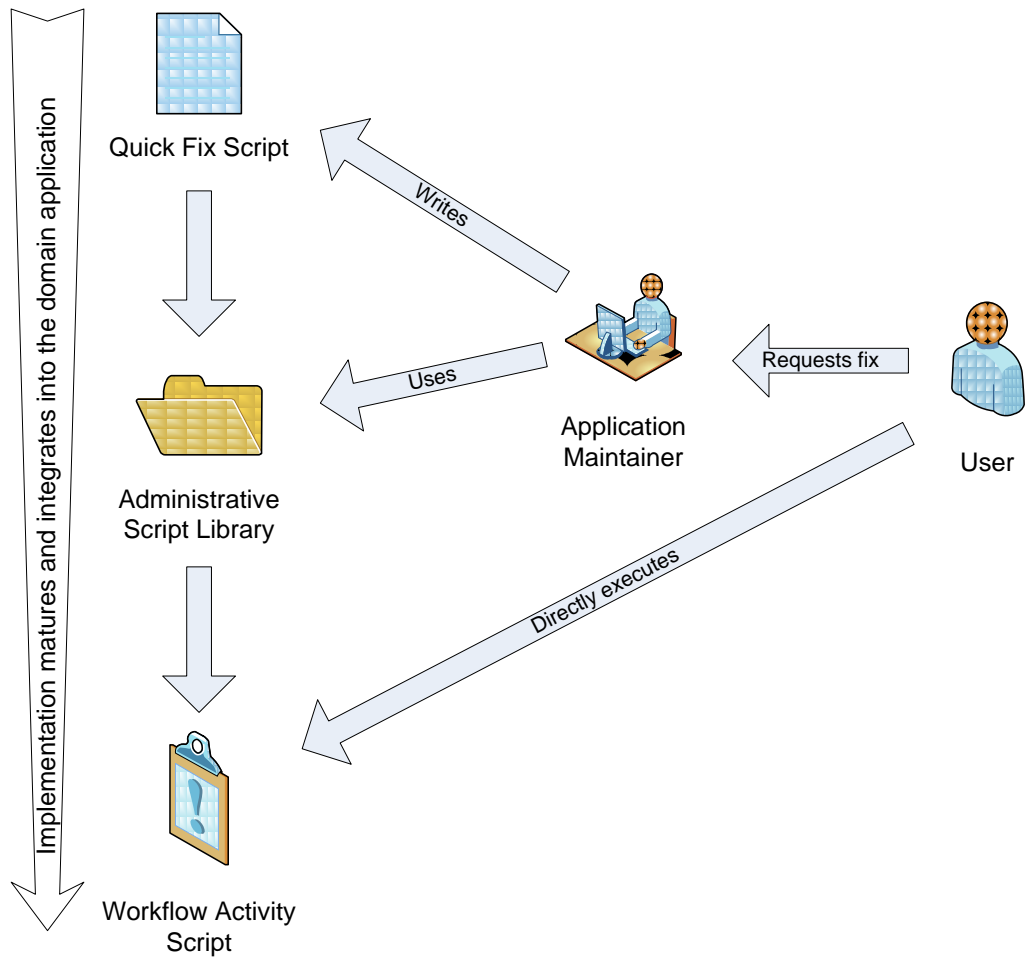

*Figure 5 - Evolution from intervention to domain logic*

It is worth noting that this is an iterative arrangement – the application maintainer in one situation is himself a user of description development tools maintained by the domain maintainer above, who deploys and maintains those tools on top of the kernel itself. To draw an analogy with the world of object orientation, the domain expert writes the classes with which (s)he then defines different types. In ECAL, the domain consists of Products and Orders, and the different detector components and the transfer requests to move them around are defined from those Orders.

*5.4 CRISTAL in Practice at the Compact Muon Solenoid ECAL*

This section outlines how the CRISTAL system performed in practice for the construction of the ECAL detectors at CERN. The CRISTAL V1.0 software was developed over the period 1997-2000 and was delivered for use at the Compact Muon Solenoid early in 2000, when the data from the characterization and the physical measurement of the lead tungstate crystals began. CRISTAL V1.0 was implemented in C++ and Java, and collected initial data against a data model residing in Objectivity but, as explained earlier, it suffered from numerous performance limitations. After redesign CRISTAL V2.0 was made available to the ECAL users from 2002 and it remained in operation over the following eight years. The latest version, on which this research is based (V3.0), is in the process of being made available as Open Source.

Each ECAL crystal generated between 2-3Mbytes of information which was mainly gathered in an automated data acquisition system, called ACCOS [15], which characterised the crystals in batches over a period of 8-10 hours for each batch of 30 crystals. Due to the vast numbers of crystals to measure it was important that this data acquisition process was reliable, consistent and that the mean time between failures was very high. Some batches required re-characterisation and some crystals were faulty and needed to be replaced; the whole data acquisition process took around five years to complete, following an initial testing period which itself took several months. It was the responsibility of one CRISTAL software engineer (the so-called application maintainer) to ensure as smooth operation as possible of the data acquisition and to provide round-the-clock accessibility to the CRISTAL database. Where changes were required to the descriptions handled by CRISTAL the procedure outlined above was followed.

During the six years of near-continuous operation, the descriptions went from beta to production then through years of (relatively few) alterations of the domain logic which necessitated very little change in the actual server software, illustrating the flexibility of the CRISTAL approach. These alterations were minor in nature and included updates to descriptions of processes and data sources which were handled by version management capability of CRISTAL. The server software only needed to be upgraded seven times, and of those seven, only one (that is the minor version change from CRISTAL V2.0 to CRISTAL V2.1 in 2004) was a required update that needed to be made available to all users and servers. This was necessary because some data formats originally designed in V2 proved not to be as scalable as required, therefore a client update was required to read the new structures.

*Table 1 - Statistics of CRISTAL operation for ECAL at CERN*

| Global ECAL CRISTAL Statistics | |
|---|---|
| Total number of centres (servers) | 9 (6 at CERN, 1 in Taiwan, 2 in Greece) |
| Runtime | August 2003 – August 2009 (6 years) |
| Total data size (at CERN) | 200GB |
| Total number of Items in one ECAL | 450,000 |
| Minor version upgrades (required client update) | 1 |
| Total number of kernel builds | 22 |
| Kernel builds requiring server software upgrade | 7 |

Given that the original release was beta, this was a remarkable achievement. The description data was flexible with regards to the kernel architecture, and was therefore portable between versions. Version 2.2 was developed with enhanced description development tool in the client by the administration team and was backwards compatible with V2.1. As the existing V2.1 servers were fully capable of deploying and using those descriptions, migration was not essential as so was not performed. Table 1 outlines some important statistics coming from the use of CRISTAL in the Compact Muon Solenoid. The statistics show the level of stability of CRISTAL across extended timescales for distributed communities of users thereby displaying the benefits of the description-driven approach in providing reliability, flexibility and ease of maintenance and ability to cope with change.

## 6. Lessons Learned with CRISTAL Usage

Certainly the main lesson learned from the CRISTAL project in coping with change was to develop a data model that had the capacity to cover multiple types of data (be they products or activities, atomic or composite in nature) and at the same time was elegant in its simplicity. To do this a disciplined and rigorously applied object-oriented approach to data modelling was required: designers needed to think in a way that would ultimately facilitate system flexibility, would enable rapid change and would ease the burden of maintenance from the outset of the design process. The approach that was followed in designing CRISTAL was to concentrate on the essential enterprise objects and descriptions (items, workflows, activities, agents, outcomes, events, viewpoints, properties and collections as described in section 3) that could be needed during the lifetime of the system no matter from which standpoint that data is accessed. Thus the system was allowed to be open in design and flexible in nature and the elegance of its design was not compromised by being viewed from one or several application-led standpoints (such as BPM, EAI, CRM, PDM or WfMS). Rather we enabled the traceability of the essential enterprise objects over the lifetime of the system as the primary goal of the system and left the application-specific views to be defined as and when they became required. The ability of description-driven systems to both cope with change and to provide traceability of such changes (i.e. the 'provenance' of the change) we see as one of the main contributions of the CRISTAL approach to building flexible and maintainable systems and we believe this makes a significant contribution to how enterprise systems can be implemented. For more detail consult paper [21] which discusses this in a practical application. Recently a start-up company called Technoledge [30] has been established to develop applications of CRISTAL that exploit this novelty.

These design skills were not simple; designers needed to be able to think conceptually, abstracting the characteristics of everyday objects into 'items' with associated metadata and to be able to represent that complexity in a concrete data model. Great benefits in terms of maintainability and flexibility resulted from being able to treat many different system objects in a single standardised manner. Savings over the lifetime of the ECAL project at CERN are estimated at several man years of effort and, as reported earlier in table 1, statistics support this assertion. The importance of instantiation and description in formulating a generic CRISTAL data model cannot be overemphasised. We propose that the description-driven design approach that emerges from this study is a genuinely new approach to designing for change.

Great importance was placed on the involvement of users at all stages of the development of CRISTAL, following many of the principles of participatory design [31]. We regard this as one of the prime reasons for the eventual success of the project. The research nature of the environment in which CRISTAL was formulated and developed led to both advantages and disadvantages. Although initially it was hoped that high-end expert users would be able to develop workflows themselves, in practice this was not possible. Instead the users collaborated closely with the designers from the outset of the project to establish a much clearer idea of the implications of their requirements, and with a full understanding of the functionality that their workflow must provide. This could then be implemented with verifiable accuracy to what the user originally specified. Essentially this approach led to a very simple way of representing new requirements and absorbing them rapidly into the evolving data model, as and when they emerged. On the negative side the users necessarily did not always know at the outset what their final requirements would be for data and process management, often leading to disruptive changes in design decisions and an evolutionary approach to prototyping. On the positive side, the users were not locked into a 'static' product: the CRISTAL model was evolving to cater for their requirements and could be made responsive to their needs. This addresses the main research question raised in this paper, that of how users can cope with changing requirements, by providing users with a means to dynamically alter the data model on-the-fly.

Control of evolving user requirements was a particularly challenging problem. New requirements needed to be addressed at the application level which, as a consequence, induced requirements further down at the domain implementation level which in turn passes its own requirements down to the kernel level. The result of this was that there could be a considerable number of potential feature configurations of the CRISTAL kernel needed to meet all possible requirements from the user. Since CRISTAL was originally conceived as an object-based system and an object-oriented approach was adopted in its design, an attempt was made to follow as far as was practically possible best software engineering practice in implementing features associated with object oriented models (e.g. inheritance, polymorphism, deferral of commitment, etc) in order to ensure reuse and extensibility. Whenever a new design modification was needed, the approach taken therefore was always to implement as open and flexible a solution as the design allowed in order not to constrain future extensions. In practice, however, this "second guessing" quickly led to feature creep and spiralling complexity which was at risk of compromising the system development process. To address this situation the approach that we adopted was to make the implementation of new requirements as intuitive as possible with as simple functionality as necessary to cope with the requirements, thereby preserving the elegance of the original (description-driven) design. This led to a closely connected set of system functionalities which was easy to maintain and to dynamically extend when required. In addition this much simpler system has the virtue of being a lot easier for users, developers and administrators new to the system to pick up and start working with.

It was also important in the implementation of CRISTAL to deliver software that would be easy to amend and update as new requirements emerged over the long timescales of the project. In fact a component-based approach to software delivery was adopted so that when new alternatives came along they could be evaluated and, where necessary, absorbed into the CRISTAL suite. Inevitably, some components were written by hand that were soon to become available as well supported open source tools (e.g. Hibernate and to a certain extent J2EE). Conversely, some third-party libraries that were expected to become *de facto* standards (e.g. Castor XML) did not prove as valuable as had been hoped, and some libraries even ended up defunct (e.g. Orbacus CORBA). However, having adopted a component-based software build, it was possible to replace modules or components on an as-needed basis fairly seamlessly.

Further evidence of the benefits accruing from use of CRISTAL comes from its commercialization as the Agilium product. Since 2004 an early version (V2.1) of the CRISTAL Kernel has been exploited by the M1i company (formerly called Agilium, based in Annecy, France) for the purpose of supporting business process management (BPM) and the integration and co-operation of multiple business processes especially in business-to-business applications. M1i have taken CRISTAL and added applications for BPM that benefit from the DDS aspects of CRISTAL, i.e. its flexibility, reusability, complexity handling and system evolution management. Their product addresses the harmonization of business processes by the use of a CRISTAL database so that multiple potentially heterogeneous processes can be integrated with each other and have their workflows tracked in the database. Agilium also integrates the management of data coming from different sources and unites BPM [32] with business activity management (BAM) [33] and Enterprise Application Integration (EAI) [34] through the capture and management of their designs in the CRISTAL system. Using the facilities for description and dynamic modification in CRISTAL, Agilium is able to provide modifiable and reconfigurable business process workflows. It uses the description-driven nature of the CRISTAL model to act dynamically on process instances already running and can thus intervene in the actual process instances during execution. These processes can be dynamically (re-) configured based on the context of execution without compiling, stopping or starting the process and the user can make modifications directly and graphically of any process parameter. Thus the Agilium system aims to provide the level of flexibility for organizations to be agile in responding to the ongoing changes required by cyber-enterprises, with functionality derived from use of CRISTAL. Details of the Agilium product and their user community can be found at [28].

## 7. Conclusions and Future Plans

The CRISTAL system has been in use at the Compact Muon Solenoid ECAL group since 2000; the production version (V2.1) of CRISTAL has been stable and has had continuous use at ECAL since 2002 and has collected over 200Gbytes of data with almost continuous performance, as outlined in the evaluation section of this paper. Without the use of CRISTAL it would not have been possible for the ECAL detector to become operational in time for first data taking at the LHC in 2011. In 2012 the ECAL detector was instrumental in the discovery of the Higgs Boson candidate at the Compact Muon Solenoid experiment at CERN [35], using the calibration data managed by CRISTAL.

The study described in this paper has demonstrated the benefits of a self-describing description-driven design approach to both designer and to users in practice. It has shown that *describing a proposed system explicitly* and openly from the outset of the project enables the developer to change aspects of it responsively as users' requirements evolve. This enables seamless transition from version to version with (virtually) uninterrupted system availability and facilitates full traceability throughout the system lifecycle. Following the principles of object-oriented design the approach encourages reuse of code, configuration data and scripts/methods. Indeed, the description-driven design approach takes this one step further and provides reuse of meta-data, design patterns [13] and maintenance of items and activities (and their descriptions). Practically this results in a higher level of control over design evolution and simpler implementation of system improvements and easier maintenance cycles. Many system elements have gained in conceptual simplicity and consequent ease of management thanks to loose typing and the adoption of a unified approach to their online manipulation: activities/scripts and their methods; member types and instances; properties and primitives; items and collections; and outcome schemas and views. One logical consequence of providing such a unified design and simplicity of management is that the CRISTAL software can be used for a wide spectrum of applications and, as a result, the applicability of the software was quickly recognized.

Future work is being directed by Technoledge [30] at upgrading CRISTAL with the ability to model domain semantics e.g. the specifics of a particular application domain e.g. healthcare, public sector, finance, and aerospace. This will essentially transform CRISTAL into a self-describing model execution engine, making it possible to build applications directly on top of the design, without code generation. The design will be the framework in which all of the application logic can be hung – without the risks of misalignment and subsequent loss that code generation can bring – and for CRISTAL to be configured as needed to support the application logic (be that BPM, EAI, CRM, PDM or WfMS). What this means is that the CRISTAL kernel will then be able to capture information about the application area in which a particular instance is being used. This will allow usage patterns to be described and captured, roles and agents to be defined on a per-application basis, and rules and outcomes specific to particular user domains to be managed. In turn this will enable multiple instances of CRISTAL to discover the semantics needed for them to inter-operate and to exchange data. This should have a profound effect in terms of business-to-business operation and the ease of configuration and maintainability of systems involving multiple instances of CRISTAL. Using this we can enable business to business (B2B) collaboration by, for example, allowing one company using CRISTAL to discover information from another company also using CRISTAL and thereby to facilitate cooperation. Using the example introduced earlier in this paper, one company supplying spark plugs in the automotive industry to another company (Engine assembly) could use the enhanced CRISTAL version to discover information it requires to enable the automatic supply of the parts e.g. by tracking versions of the parts that are required and used in Newcar models (to use the example from section 4, above). This would consequently provide B2B functionality at the production level and facilitate inter-company collaboration.

Another emergent quality of CRISTAL is as a migratory tool to integrate legacy systems. Its ability to run more than one model concurrently and the flexibility of its model means that it can be used to emulate multiple existing systems simultaneously in order to integrate them into a common environment. A new unified model can them be implemented alongside all of the old data, supporting it as much as is needed. Once there is no further need for the old models, the new one can then be migrated out into a new system if and when required. In the ECAL system at CERN, a small custom server module was implemented to provide a simple telnet interface for machines controlled with LabVIEW to connect to CRISTAL. The LabVIEW programs required few changes to submit their data – little beyond changing the connection parameters.

Recently CRISTAL has been studied as a 'scientific provenance data management system' i.e. a system that is used to manage the history of data and workflows and their usage over time in a scientific environment [20]. The neuGRID project supported by the EC 7th Framework Programme is implementing Provenance and Querying Services using a CRISTAL kernel for the purposes of supporting neuroscientists in their studies of Alzheimer's disease across Europe (see [36] and [37]). The use of CRISTAL as the Provenance Service database and engine has enabled neuroscientists to track their complex image analyses over time and to collaborate together in teams with fully versioned workflows and datasets. Furthermore in the context of provenance data management, the CRISTAL data model is currently being adapted for compliance with the emerging standard Open Provenance Model (OPM, [38]) in the so-called Virtual Laboratory environment of the follow-up project to neuGRID, called N4U [39]. In N4U CRISTAL acts both as a provenance tracking system and as a knowledge base for researchers to consult in building and conducting their analyses.

Research into the further extension and uses of CRISTAL continues. There are plans to enrich its kernel (the data model) to model not only data and processes (products and activities as items) but also to model agents and users of the system (whether human or computational). It is planned to investigate how the semantics of CRISTAL items and agents could be captured in terms of ontologies and thus mapped onto or merged with existing ontologies for the benefit of new domain models. The emerging technology of cloud computing and its application in complex domains, such as medicine and healthcare, provide further interesting challenges, particularly in healthcare [40].

CRISTAL will be used in further production management techniques both at CERN and other research environments. The Technoledge company is engaging with a French research institution to use CRISTAL to track all the production processes in the design and development of fuel cells for the next generation of electric vehicles. It is also working with firms in the accountancy domain to use CRISTAL to integrate multiple legacy systems and to build a common ongoing development environment. Furthermore at the University of the West of England it is planned that CRISTAL will be used to track the building of a modern sports and leisure complex, gathering data from architects, construction managers, engineers and built environment experts in order to provide both support for the construction process and to act as a source of acquired educational knowledge for students across that university [41]. It is the nature of the 'open' and flexible description-driven model that underpins CRISTAL, as outlined in this paper, that has allowed these diverse studies to take place.

## Acknowledgements

The authors wish to thank their respective institutions for the support given during the development of the CRISTAL software. The long-scale development, testing and refinement of the CRISTAL design philosophy and model structure would not have been possible without the continued support of those institutions. In particular they would like to acknowledge the invaluable help and advice given to them by the following : Jean-Pierre Vialle, Thierry Le Flour, Sophie Lieunard, Alain Bazan and Steve Murray then of LAPP/CNRS, Annecy (France), Etiennette Auffray, Paul Lecoq and Guy Chevenier from CERN (Switzerland) and Zsolt Kovacs, Tony Solomonides, Florida Estrella and Norbert Toth from the University of the West of England (UK). They would also like to acknowledge the positive contributions from a number of external organisations that have allowed the exploitation of CRISTAL in practical commercial and research environments.